# Polarization-selective vortex-core switching by orthogonal Gaussian-pulse currents

Young-Sang Yu,[1] Ki-Suk Lee,[1] Hyunsung Jung,[1] Youn-Seok Choi,[1] Myoung-Woo Yoo,[1] Dong-Soo Han,[1] Mi-Young Im,[2] Peter Fischer,[2] and Sang-Koog Kim[1,2*]

[1]*National Creative Research Initiative Center for Spin Dynamics & Spin-Wave Devices, and Nanospinics Laboratory, Research Institute of Advanced Materials, Department of Materials Science and Engineering, Seoul National University, Seoul 151-744, South Korea*

[2]*Center for X-ray Optics, Lawrence Berkeley National Laboratory, Berkeley CA 94720, USA*

We experimentally demonstrate low-power-consumption vortex-core switching in magnetic nanodisks using tailored rotating magnetic fields that are produced with orthogonal and unipolar Gaussian-pulse currents. Optimal width of the orthogonal pulses and their time delay are found to be determined only by the angular eigenfrequency $\omega_D$ for a given vortex-state disk of its polarization $p$, such that $\sigma = 1/\omega_D$ and $\Delta t = \frac{\pi}{2} p/\omega_D$, as studied from analytical and micromagnetic numerical calculations. The estimated optimal pulse parameters are in good agreements with the experimentally found results. This work provides a foundation for energy-efficient information recording in vortex-core cross-point architecture.

## I. INTRODUCTION

Magnetic vortices in patterned magnetic dots are a very promising candidate for information storage applications, not only because of the energetically stable twofold ground state of their core magnetization[1-5] but also due to a reliable core switching dynamics explored so far.[6-10] A lot of experimental, analytical and micromagnetic numerical studies have revealed that low-power driven vortex-core switching in magnetic nanodots can be simply achieved by applications of not only in-plane alternating[6] or rotating magnetic fields[11] or spin-polarized currents,[12] but also their pulse forms.[7,8,13-15] Ever since, the fundamentals of ultrafast vortex-core switching dynamics, including their mechanism,[6-10] physical origin,[12,16] and criterion[10,12,17-19] have been subjects of intense research. However, an important but very challenging issue of low-power information-writing using orthogonally crossed electrodes that are available in the basic cross-point architecture has not been demonstrated yet.

In this article, we experimentally demonstrated polarization-selective vortex-core switching using rotating magnetic fields, achieved using orthogonal and unipolar Gaussian-pulse currents. In order to achieve energy-efficient switching using those simple electrodes, we optimized analytically and numerically the width of the orthogonal pulses and their time delay based on Thiele's analytical approach. Using the optimal values, we successfully demonstrated low-power driven vortex-core switching in experiments.

## II. CONCEPT OF IDEAL AND PULSE-TYPE ROTATING FIELDS

In one of our earlier works, we proposed the unique novel means of circular-rotating magnetic fields for an efficient manipulation of polarization-selective vortex-core reversals.[5,18,20] Ideal single harmonic circular-rotating fields can be expressed as the vector sum of two linearly oscillating harmonic fields as shown in Fig. 1 and, mathematically, as $\mathbf{H}_{CCW} = H_0 \cos(\omega_\mathbf{H} t)\hat{\mathbf{x}} + H_0 \sin(\omega_\mathbf{H} t)\hat{\mathbf{y}}$ and $\mathbf{H}_{CW} = H_0 \cos(\omega_\mathbf{H} t)\hat{\mathbf{x}} - H_0 \sin(\omega_\mathbf{H} t)\hat{\mathbf{y}}$; Such a type of magnetic field rotates either counter-clockwise (CCW) or clockwise (CW) on the film plane with a constant magnetic field strength $H_0$ and a single-harmonic frequency $\omega_\mathbf{H}$ (see Fig. 1). It was reported that the CCW (CW) rotating field can switch only the upward (downward) core to its reversed orientation with the lowest $H_0$ value when the field frequency $\omega_\mathbf{H}$ is tuned to the vortex angular eigenfrequency $\omega_D$.[5,11,18] However, such ideal rotating fields are not readily available in real devices.

Our approach in the present work is thus to use, for efficient vortex-core switching, tailored special fields achieved with unipolar Gaussian-pulse currents flowing along two orthogonal electrodes;[11] a cross-type electrode comprised of two striplines orthogonally arranged, as shown in Fig. 2(a). The simplicity of this design builds upon already existing actual memory devices based on the cross-point architecture. The Gaussian-shape current pulses can be characterized simply by the pulse width $\sigma$ and the height $I_0$. Oersted fields are generated around both current paths. The local fields at the cross-point of the two orthogonal electrodes vary in their strength and direction with time and space on the film plane, yielding certain types of rotating field. The rotation sense of the resultant rotating fields varies primarily with delay time $\Delta t$ between the two orthogonal currents [see Figs. 2(b) and 2(c)]. For example, $\Delta t = 0$ indicates certain linear fields, but $\Delta t < 0$ and $\Delta t > 0$ represent the CW and CCW sense of the rotating fields, respectively. The dominant frequency can be determined by $\sigma$. Thus, a certain combination of the $\Delta t$ and $\sigma$ values represents a special rotating field, the so-called pulse-type rotating field in order to distinguish it from ideal single harmonic rotation fields mentioned earlier.

### III. OPTIMIZING THE CURRENT PULSE PARAMETERS

#### A. Analytical derivations

In order to find the optimal values of the pulse parameters by which vortex-core switching events can be achieved by the lowest value of $H_0$, we employed analytical derivations and numerical calculations of the vortex-core motions in response to different pulse parameters, based on Thiele's equation.[21]

Rotating magnetic fields obtained from the aforementioned orthogonal unipolar Gaussian-pulse currents can be expressed mathematically as $\mathbf{H}_{Rot} = \mathbf{H}_x + \mathbf{H}_y$ with $\mathbf{H}_x = H_{x0} \exp(-t^2/(2\sigma^2))\hat{\mathbf{x}}$ and $\mathbf{H}_y = H_{y0} \exp(-(t-\Delta t)^2/(2\sigma^2))\hat{\mathbf{y}}$ as seen in Fig. 3(a). Such pulse-type rotating fields, in contrast to the corresponding ideal circular-rotating fields, have time- and space-varying field strengths [see Fig. 2(c)] as well as many different harmonic frequencies along with a certain frequency profile. This type of field can be considered to be a mixture of ideal circular-rotating fields, each of which has a different constant field strength and different single harmonic frequency, as shown in Figs. 3(b) and 3(c). In reference to the circular-rotating-frame basis and complex variables, an ideal circular-rotating field of strength $h_0$, phase $\phi$, and frequency $\omega_h$ can be expressed mathematically as $\tilde{h}\,e^{i\omega_h t} = h_0\,e^{i\phi}\,e^{i\omega_h t} = h_0 \cos(\omega_h t + \phi) + i h_0 \sin(\omega_h t + \phi)$. Here $\omega_h > 0$ and $\omega_h < 0$ correspond to the CCW and CW rotation sense, respectively. The power spectra of individual harmonics vary markedly with the values of $\Delta t$ and $\sigma$, as shown in Figs. 3(b) and 3(c), respectively, as calculated from $\tilde{h}(\omega_h) = \frac{1}{\sqrt{2\pi}} \int_{-\infty}^{\infty} \tilde{H}(t) e^{-i\omega_h t} dt$ with $\tilde{H} = H_{x0} \exp(-t^2/(2\sigma^2)) + iH_{y0} \exp(-(t-\Delta t)^2/(2\sigma^2))$ and $H_{x0} = H_{y0} = H_0$.

To derive explicit analytical equations, we start with a susceptibility tensor[20] for vortex-core motions driven by linearly polarized harmonic fields: $\hat{\chi}_{X,L}(\omega_\mathbf{H}) = \mathbf{X}_{0,L}/\mathbf{H}_{0,L}$, with the vortex-core position vector $\mathbf{X} = \mathbf{X}_{0,L} \exp(-i\omega_\mathbf{H} t)$, the harmonic oscillating field $\mathbf{H} = \mathbf{H}_{0,L} \exp(-i\omega_\mathbf{H} t)$ and the angular frequency $\omega_\mathbf{H}$ ( > 0). The equation of $\hat{\chi}_{X,L}(\omega_\mathbf{H}) = \mathbf{X}_{0,L}/\mathbf{H}_{0,L}$ can be rewritten, in matrix form, as

$$\begin{pmatrix} X_{0,x} \\ X_{0,y} \end{pmatrix} = \frac{\mu}{(i\omega_\mathbf{H} D + \kappa)^2 - (\omega_\mathbf{H} G)^2} \begin{bmatrix} -i\omega_\mathbf{H} G & -i\omega_\mathbf{H} D - \kappa \\ i\omega_\mathbf{H} D + \kappa & -i\omega_\mathbf{H} G \end{bmatrix} \begin{pmatrix} H_{0,x} \\ H_{0,y} \end{pmatrix}, \qquad (1)$$

where $\mathbf{X}_{0,L} = X_{0x}\hat{\mathbf{x}} + X_{0y}\hat{\mathbf{y}}$ and $\mathbf{H}_{0,L} = H_{0x}\hat{\mathbf{x}} + H_{0y}\hat{\mathbf{y}}$, along with $\mu = \pi R L M_s \xi C$ for the circular disk with radius $R$, thickness $L$, and the saturation magnetization $M_s$, the chirality of vortex $C$ ($C = +1$ for the CCW in-plane curling magnetization direction and $C = -1$ for the CW direction), and $\xi = 2/3$ for the "side-charge-free" model, the stiffness coefficient $\kappa$, the gyrovector constant $G$, and the damping constant $D$.[22] The vortex-core displacement from the center position in response to a single-harmonic rotating field, $\tilde{h} e^{i\omega_h t}$ can thus be expressed simply as $\tilde{X} = \chi_{\omega_h} \tilde{h}$, with the complex-variable position vector $\tilde{X} e^{i\omega_h t} \equiv X_x + iX_y$, the single-harmonic circular-rotating field $\tilde{h} e^{i\omega_h t} = h_x + ih_y$ and the complex susceptibility $\chi_{\omega_h} = i\mu/(\kappa - \omega_h G - i\omega_h D)$.[18] Finally, integrating all of the vortex-core motion responses to individual harmonic fields, expressed mathematically as $\tilde{X}(t) = \frac{1}{\sqrt{2\pi}} \int_{-\infty}^{\infty} \chi_{\omega_h} \tilde{h}(\omega_h) e^{i\omega_h t} d\omega_h$, gives rise to resultant positions of dynamic vortex-core motion.

Figure 4 shows the results of the numerical calculations of $|\chi_{\omega_h}|$ and $|\chi_{\omega_h} \tilde{h}(\omega_h)|$ as a function of the reduced parameter of $\omega_h/\omega_d$ for the indicated $\Delta t/\sigma$ values. The absolute value of $\tilde{X}(\omega_h) = \chi_{\omega_h} \tilde{h}(\omega_h)$ represents the steady-state radius of the corresponding vortex-core displacement at each harmonic frequency. Given the fact that vortex-core gyrations arise resonantly at $\omega_h = \omega_d$ (where $\omega_d = p\omega_D$), the gyration-assisted low-power-driven switching can be maximized with $\omega_h = \omega_d$, where $\omega_D = \kappa|G|/(G^2 + D^2)$ with the same notations defined in Ref. 20. The maximum value of $|\tilde{h}(\omega_h)|$ can be found at $\omega_h = \omega_d$, where $\tilde{h}(\omega_h) = H_0 \sigma \exp(-\sigma^2 \omega_h^2/2)(1 + i\exp(-i\omega_h \Delta t))$, and is given as $|\tilde{h}(\omega_d)| = H_0 \sigma \exp(-\sigma^2 \omega_d^2/2)\sqrt{2 + 2\sin(\omega_d \Delta t)}$. The $|\tilde{h}(\omega_d)|$ at $\omega_h = \omega_d$ varies dramatically with $\Delta t$ at a constant value of $\sigma = 1/\omega_D$ and with $\sigma$ at a constant value of $\Delta t = \frac{\pi}{2} p/\omega_D$, as shown in Figs. 3(b) and 3(c), respectively. The numerical calculations of $|\tilde{h}(\omega_d)|/H_0 = \sigma \exp(-\sigma^2 \omega_d^2/2)\sqrt{2 + 2\sin(\omega_d \Delta t)}$ show the maxima of $|\tilde{h}(\omega_d)|$ at

$\sigma = 1/\omega_D$ and $\Delta t = p(2n+\tfrac{1}{2})\pi/\omega_D$ with $n$ = 0, ±1, ±2, … (see the dashed vertical gray lines in Fig. 5). Note that the optimal pulse parameters are determined only by $\omega_D$ for a given vortex of its polarization $p$. We also emphasize that the dominant frequencies (peaks) in the power spectra of harmonics are not equal to the eigenfrequency at which $\sigma$ and $\Delta t$ are optimized to obtain the largest value of $|\tilde{h}(\omega_h = \omega_d)|$.

### B. Micromagnetic numerical calculation

The above analytical estimates were also confirmed from micromagnetic simulations for vortex-core motions, responding to different values of $\sigma$ and $\Delta t$, in a Py disk of radius $R$ = 150 nm and thickness $L$ = 20 nm, for the upward core orientation ($p$ = +1). Here, to reduce the computation time, we used a much smaller Py disk than will be used in the experiment. We employed the OOMMF code,[23] which utilizes the Landau-Lifshitz-Gilbert equation of motion of magnetization, $\partial \mathbf{M}/\partial t = -\gamma(\mathbf{M}\times\mathbf{H}_{\text{eff}}) + \alpha/|\mathbf{M}|(\mathbf{M}\times\partial\mathbf{M}/\partial t)$ [Ref. 24], with the gyromagnetic ratio $\gamma$ (2.21 × 10$^5$ m/As) and the phenomenological damping constant $\alpha$ (0.01). We set the unit cell size at 2 × 2 × 20 nm$^3$. The standard Py material parameters used are as follows: the saturation magnetization $M_s$ = 8.6 × 10$^5$ A/m, the exchange stiffness $A_{\text{ex}}$ = 1.3 × 10$^{-11}$ J/m, and the magnetocrystalline anisotropy constant $K$

= 0. The simulation results (symbols) of the maximum radius of the vortex-core gyrations versus $\sigma$ (at $\Delta t = 0$) and $\Delta t$ (at $\sigma = 1/\omega_D$) are shown in Figs. 6(a) and 6(b), respectively. The solid lines indicate the numerical solution of the analytically derived equations. The simulation and analytical calculations agree well quantitatively. The values of $\sigma$ = 274.4 ps and $\Delta t$ = 431.0 ps, which were obtained by inserting $\omega_D = 2\pi \times 580$ MHz into the analytically forms of $\sigma = 1/\omega_D$ and $\Delta t = \frac{1}{2}\pi/\omega_D$, are very close to the simulation values of $\sigma$ = 275 ps and $\Delta t$ = 431 ps.

## IV. EXPERIMENTAL OBSERVATION

### A. Sample preparation

As a next step, we experimentally investigated vortex-core switching events driven by pulse-type rotating fields, using a sample consisting of crossed electrode and Permalloy (Py: $Ni_{80}Fe_{20}$) disk on it, as illustrated in Fig. 2(a). The Py films were deposited by magnetron sputtering under base pressures of less than $5\times10^{-9}$ Torr. Py disks of radius $R$=2.5 $\mu$m and thickness $L$=70 nm were patterned by typical e-beam lithography (Jeol, JBX9300FS) and subsequent lift-off processes. Each disk was placed at the intersection of two 50-nm-thick stripline Au electrodes and capped with 2-nm-thick Pd layers (without breaking the vacuum) to prevent oxidation. Such cross-type electrodes similar to that employed in Ref. 11 were patterned by e-beam lithography and subsequent lift-off processes. In order to obtain sufficient soft x-ray transmission through the sample, the electrodes were deposited onto the 200-nm-thick silicon nitride membranes of a 5 mm-by-5 mm window by electron-beam evaporation under base pressures of less than $1\times10^{-8}$ Torr.

### B. Soft X-ray microscopy measurements

Vortex-core switching events were then examined involving the aforementioned pulse-type rotating fields of different values of $I_0$ ($H_0$), $\sigma$, and $\Delta t$. In the present study, the strength $H_0$ of the fields was estimated directly from $I_0$ through Ampere's law. By varying the pulse-parameter values, we manipulated the polarization, strength and frequency profiles of those fields as shown in Fig. 2(c). In order to directly monitor the out-of-plane magnetizations of the initial core orientations, either upward or downward, we employed high-resolution magnetic transmission soft x-ray microscopy (MTXM) at the Advanced Light Source in Berkeley, California [Fig. 7(a)], which utilizes the x-ray magnetic circular dichroism (XMCD) contrast at the Fe $L_3$ edge.[25] A scanning electron microscope image of a single Py disk and the representative MTXM images around the core region are shown in Fig. 7(b). The dark and white spots indicate the upward and downward core orientations, respectively, in our experimental setup, before and after a vortex-core switching event. By monitoring the dark and white spots in these experimental XMCD images, we can readily determine whether vortex-core switching events had occurred for a given field pulse.

## V. RESULTS AND DISCUSSION

We present some experimental results along with the prediction of the boundary diagrams of vortex-core switching events driven by pulse-type, CCW rotating fields and linear fields. Figure 8 shows the results (open and closed symbols) of three different Py disks of the same $R = 2.5$ $\mu$m and $L = 70$ nm dimensions at several different values of $\sigma$ for the indicated field polarizations, $\Delta t = +2.0$ ns (CCW rotating field) and $\Delta t = 0$ (linear field). In the experiment, the upward-to-downward core switching by the CW rotating fields could not been measured, because much higher field strengths are required for the switching with this type of rotating field, as shown by the orange curve. Analytical expressions of the minimum field strengths of the CCW (purple) and CW (orange) rotating fields and linear (green) fields were derived on the basis of the linearized Thiele's equation[21] of vortex-core motions and the formulated vortex-core critical velocity.[18] As is apparent in Fig. 8, the analytically calculated boundaries (solid lines) were in general agreement with the experimental ones. The analytically derived $\sigma$ value (vertical line at $\sigma$ = 1.09 ns) is close to the experimental value, $\sigma$ = 1.27 ns. This fact seemingly is associated with the difference in the eigenfrequency between the real sample and the model we used for the analytical calculation ($\omega_D/2\pi = 146$ MHz).

Next, we investigated vortex core switching driven by different values of $\Delta t$ at the constant optimal value of $\sigma$ = 1.27 ns obtained from the above experiment result. Figure 9(a) shows the experimentally found boundary curves (symbols), the minimum values of $I_0$ ($H_0$), which are the so-called threshold values required to switch the initial vortex core of upward (downward) magnetization by the CCW (CW) polarization of pulse-type rotating fields. Each polarization, $\Delta t$ = 0, ±1, ±2, and ±3 ns, is illustrated in the inset. For the case of $\Delta t$ = +2.0 ns (CCW rotating field), the upward core maintained its initial orientation with $H_0$ less than 11.0 Oe (indicated by open symbols), and then switched to the downward core orientation with $H_0 \geq$ 11.0 Oe (indicated by solid symbols) [Fig. 9(b), right column]. Even with a further increase to $H_0$ = 17.2 Oe, the once-reversed downward core was not switched back to the initial upward orientation by the CCW rotating field, because the field strengths are too much small for that field to reverse the downward core. By contrast, for the other polarization, $\Delta t = -2.0$ ns (CW rotating field), only the downward core was switched to the upward orientation with $H_0 \geq 11.0$ Oe [Fig. 9(b), left column]. For the other rotating fields (i.e., $\Delta t = \pm 1$ and $\pm 3$ ns) the same switching behaviours were observed, but with relatively high fields, because they were farther away from the corresponding ideal circular fields than the fields of $\Delta t = \pm 2.0$ ns. In the case of the linear fields, that is, $\Delta t = 0$, field strengths larger than $H_0 = 14.0$ Oe were required to switch either the upward or downward core orientation. The threshold strength of the linear fields was higher than that of the CCW (CW) rotating fields (11.0 Oe) for the upward (downward) vortex-core switching. The actual threshold field strength for this linear field is the vector-sum of both fields generated by the two orthogonal striplines at their interaction, that is, $\sqrt{2}H_0 = 19.8$ Oe. Our experimental results confirmed completely the analytical predictions of the upward (downward)-to-downward (upward) core switching at the given $\Delta t$ values [the solid lines in Fig. 9(a)]. The analytical estimates also were confirmed to be in good agreement with simulation results obtained using the same rotating fields but smaller dimensions of Py disk

We want to emphasize the following significant points. The threshold field strength required for switching a vortex core of a given polarization varies significantly with $\Delta t$, which is to say, the polarization of applied pulse-type rotating fields. The minimum threshold values were obtained at $\Delta t = +2.0$ (–2.0) ns for the upward (downward) core switching. Further, it should be noted that we achieved vortex-polarization selective switching by choosing either the CCW or CW rotating field even using unipolar Gaussian-pulse currents by changing the sign of the delay time $\Delta t$, and that a significant reduction of the core-switching field was achieved simply by using the optimal pulse parameters of $\sigma = 1/\omega_D$ and $\Delta t = \frac{1}{2}\pi/\omega_D$, rather than using arbitrary pulse parameters.

## VI. POSSIBLE MECHANISM OF VORRTEX-CORE READOUT

An additional important remark drawn from the above experimental demonstration is that, similar to the asymmetric switching response of the upward or downward core to the CCW and CW rotating fields, there are large asymmetric vortex-core-gyration amplitudes, i.e., resonantly excited versus not-excited motions, depending on the relative rotation sense of the employed rotating fields with respect to a given core magnetization. Figure 10 shows the simulation results of the trajectories of vortex-core gyrations of $p = +1$ and $p = -1$ in response to the pulse-type CCW and CW rotating fields. As consistent with the asymmetry of the switching behaviours evidenced in Fig. 9, the largest asymmetry in the vortex-core gyration amplitude between $p \cdot \text{sign}(\Delta t) = +1$ and $-1$ is obtained at $|\Delta t| = \frac{1}{2}\pi/\omega_D$. For the case of $p = +1$, only the CCW field ($\Delta t = +\frac{1}{2}\pi/\omega_D$) effects such a large-amplitude vortex-core gyration, but the CW field ($\Delta t = -\frac{1}{2}\pi/\omega_D$) does not. A large-amplitude vortex-core gyration yields a large deviation of in-plane curling magnetization from the ground state vortex, which in turn gives rise to the corresponding difference in magnetizations between the free and fixed vortex layers (e.g., in spin value structure). Provided that the rotation sense of the fields applied as the input signal is known, readouts of either core orientation by reference to low and high tunnelling magnetoresistance (TMR) through the magnetic tunnelling junction (MTJ) is possible.[26] Thus, given a typical spin valve structure that includes a MTJ, the change in TMR can be used as readout means for detecting either vortex-core orientation.

## VII. CONCLUSION

The low-power-consumption switching of vortex cores in nanodisks by two orthogonal addressing electrodes demonstrated in the present work would be a promisingly robust mechanism for information writing operation in a vortex-state dot array. The present experimental demonstration of vortex-core switching, along with the proposed possible readout mechanism, is promising for non-volatile information recording, writing, and readout operations. Thus, this work imparts further momentum to the realization of vortex random access memory based on the unique vortex structure and its novel dynamic properties.

## ACKNOWLEDGEMENTS

This research was supported by the Basic Science Research Program through the National Research Foundation of Korea (NRF) funded by the Ministry of Education, Science, and Technology (Grant No. 20100000706). S.-K.K. was supported by the LG YONAM foundation under the Professors' Overseas Research Program. Use of the soft X-ray microscope was supported by the Director, Office of Science, Office of Basic Energy Sciences, Materials Sciences and Engineering Division, U.S. Department of Energy.

**Figure captions**

FIG. 1. (Color online) (a) Schematic illustration of ideal counter-clockwise (CCW) and clockwise (CW) circular-rotating magnetic fields. (b) Time-varying oscillating magnetic fields produced by two orthogonal alternating currents.

FIG. 2. (Color online) (a) Schematic illustration of sample consisting of single Py disk and two-crossed-stripline electrode. The color and height of the indicated initial vortex state indicate the local in-plane and out-of-plane magnetization components, respectively. (b) Two Gaussian-pulse currents flowing along $x$ (blue) and $y$ (red) axes with indicated different $\Delta t/\sigma$ values. (c) Pulse-type rotating fields at intersection of two striplines. The dotted arrows indicate the rotation sense of the resultant fields on the plane.

FIG.3. (Color online) (a) Two pulse fields produced by the corresponding Gaussian-pulse currents, along with given $\Delta t$ and $\sigma$ values. Power spectra versus harmonic frequency for pulse-type rotating fields resulting from the vector sum of two orthogonal fields for indicated different values of $\Delta t/\sigma$ (at $\sigma =1/\omega_D$) in (b) and $\sigma \cdot \omega_D$ (at $\Delta t \cdot \omega_d =+\pi /2$) in (c), where $\omega_d/2\pi = 580$ MHz.

FIG. 4. (Color online) Calculations of (a) $|\chi_{\omega_h}|$ and (b) $|\chi_{\omega_h}\tilde{h}(\omega_h)|$ versus $\omega_h/\omega_d$ for up-core gyration, in Py disk of $R = 150$ nm and $L = 20$ nm, by pulse-type rotating fields of $\sigma =$ 274.4 ps and indicated different values of $\Delta t/\sigma$.

FIG. 5. (Color online) $|\tilde{h}(\omega_d)|/H_0$ versus $\sigma \cdot \omega_D$ at $\Delta t = 0$ in (a) and versus $\Delta t \cdot \omega_d$ at $\sigma =1/\omega_D$ in (b).

FIG. 6. (Color online) Comparison between analytical calculation (solid lines) and micromagnetic simulation (symbols) for the maximum radius $|X|_{max}$ of gyrotropic motions of upward core orientation ($p = +1$, $\omega_d =\omega_D$) in Py disk of $R = 150$ nm and $L = 20$ nm. $H_0 = 50$ Oe was used in both the simulation and analytical calculations.

FIG. 7. (Color online) (a) Schematic illustration of our experimental setup for MTXM measurements. The two Fresnel zone plates project a full-field image of the spatial

distribution of the local magnetizations onto a soft-x-ray-sensitive CCD camera. In the experiment, the soft x-rays were incident normal to the surface of the samples so that the out-of-plane components of the vortex-core magnetizations could be directly monitored. (b) Left: Scanning electron microscope image of single Py disk of radius $R$ = 2.5 $\mu$m and thickness $L$ = 70 nm, placed on two-crossed-stripline Au electrode. The spatial distribution of the local out-of-plane magnetizations around the core regions, for upward (middle) and downward (right) core orientation, before and after switching.

FIG. 8. (Color online) Boundary diagram of vortex-core switching events on plane of $\sigma$ and $I_0$ ($H_0$). The symbols and solid lines indicate the experimental results and analytical calculations, respectively. The experimental results were obtained from three different samples of identical dimensions, as indicated by I, II, and III. The closed and open symbols indicate the occurrence of vortex-core switching and non-switching events, respectively. The solid vertical line indicates the optimal value of the analytical calculation ($\sigma$ = 1.09 ns).

FIG. 9. (Color online) Vortex-core switching diagram on $\Delta t$ - $I_0$ ($H_0$) plane. (a) Boundary diagram of vortex-core switching events, experimentally observed with applications of two orthogonal and unipolar Gaussian-pulse currents of varying $I_0$ ($H_0$) at different polarizations $\Delta t = 0, \pm 1, \pm 2$, and $\pm 3$ ns, with the optimized value of $\sigma = 1.27$ ns. The vertical dashed lines indicate the various polarizations of the linear and several rotating fields represented by the shapes shown below. The closed and open symbols denote the core switching and non-switching events, respectively, at the corresponding field strengths and types. The purple and orange solid lines indicate the analytical calculation of the threshold field strengths versus $\Delta t$ for the up-to-down and down-to-up core switching, respectively. (b) Zoomed illustrations of XMCD contrast images at core regions before and after vortex-core switching driven by indicated types of applied fields.

FIG. 10. (Color online) Asymmetry of vortex-core gyration amplitudes, represented by trajectories of upward and downward cores driven by pulse-type rotating fields with $\Delta t = \pm \frac{1}{2} \pi / \omega_D$, where $\omega_D / 2\pi = 580$ MHz for Py disk of $R = 150$ nm and $L = 20$ nm. The field strength used, $H_0$ = 150 Oe, was weaker than the threshold field strength for vortex-core switching in the given disk dimensions.

**FIG. 1.**

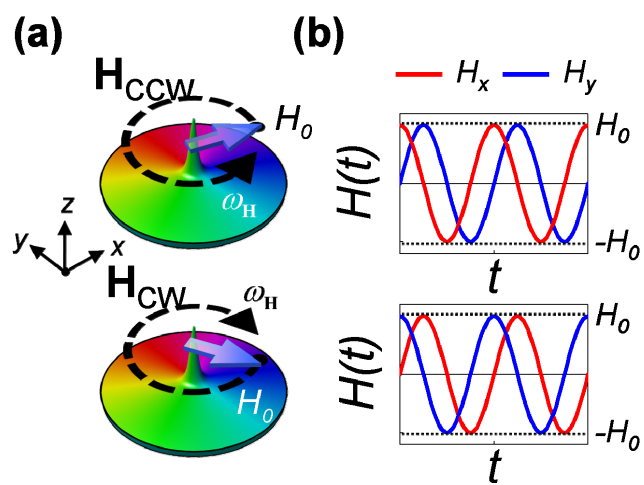

**FIG. 2.**

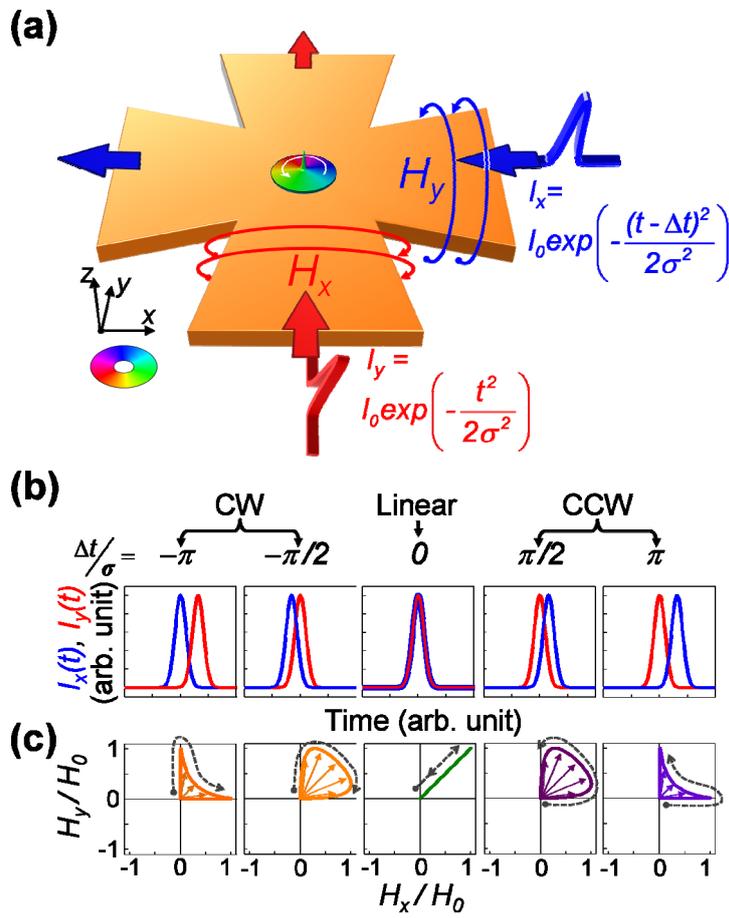

**FIG. 3.**

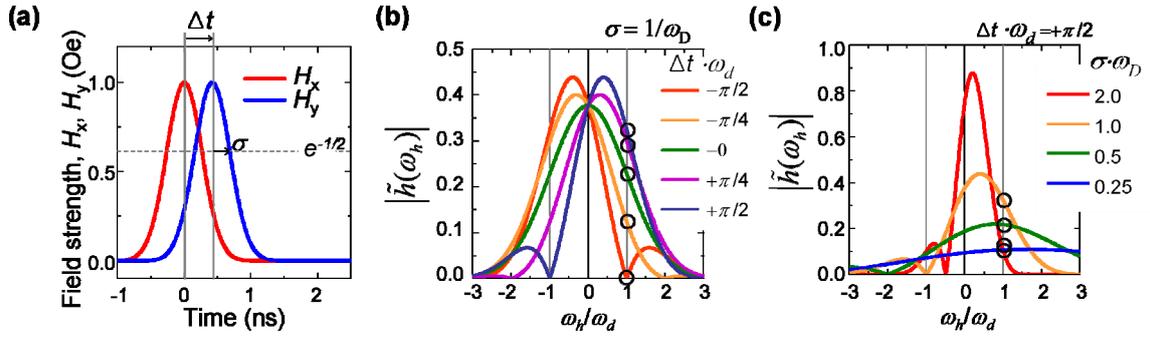

**FIG. 4.**

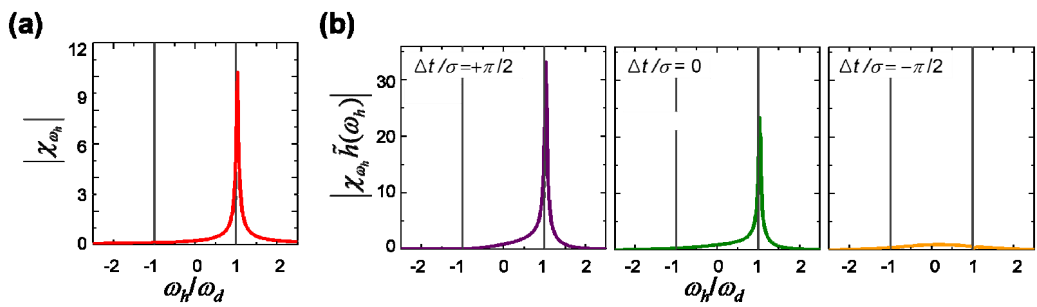

**FIG. 5.**

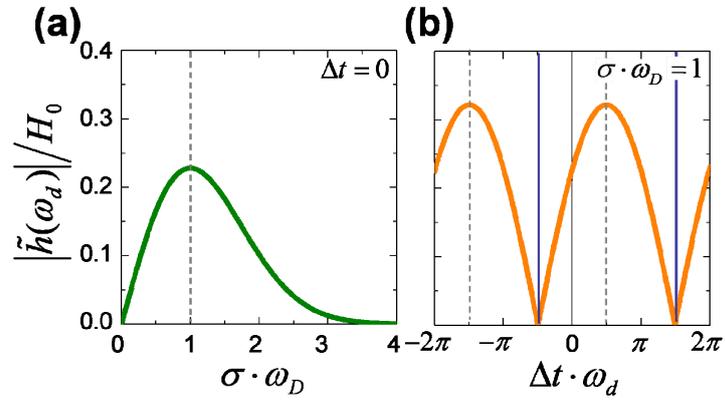

**FIG. 6.**

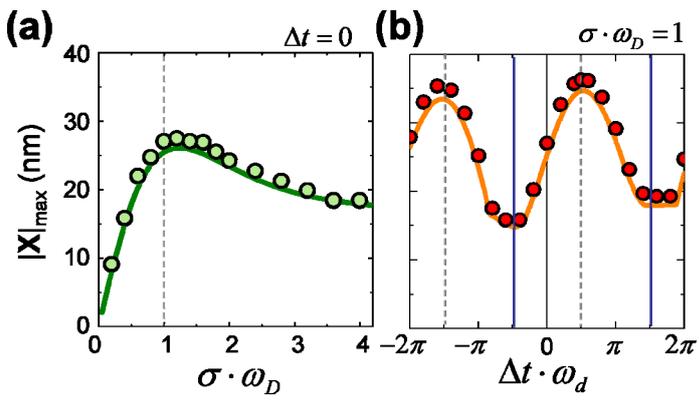

**FIG. 7.**

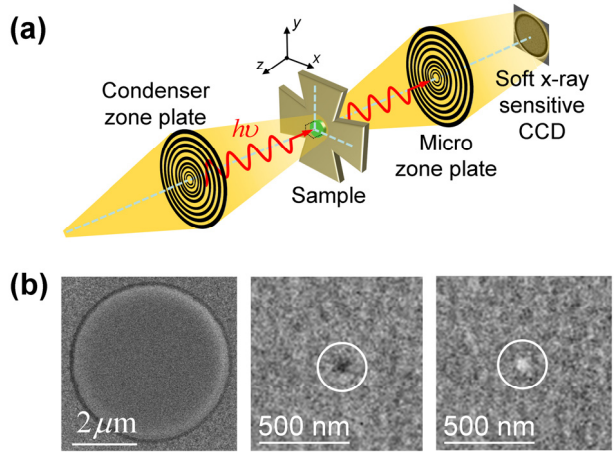

**FIG. 8.**

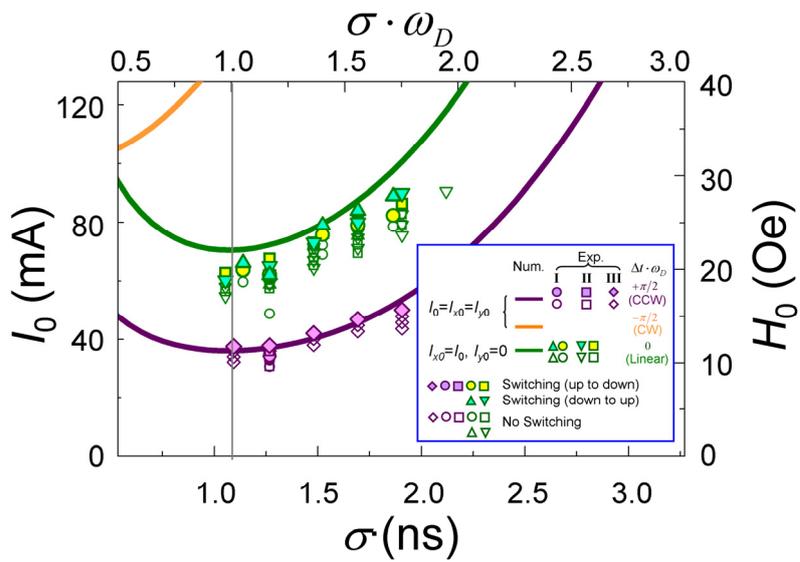

**FIG. 9.**

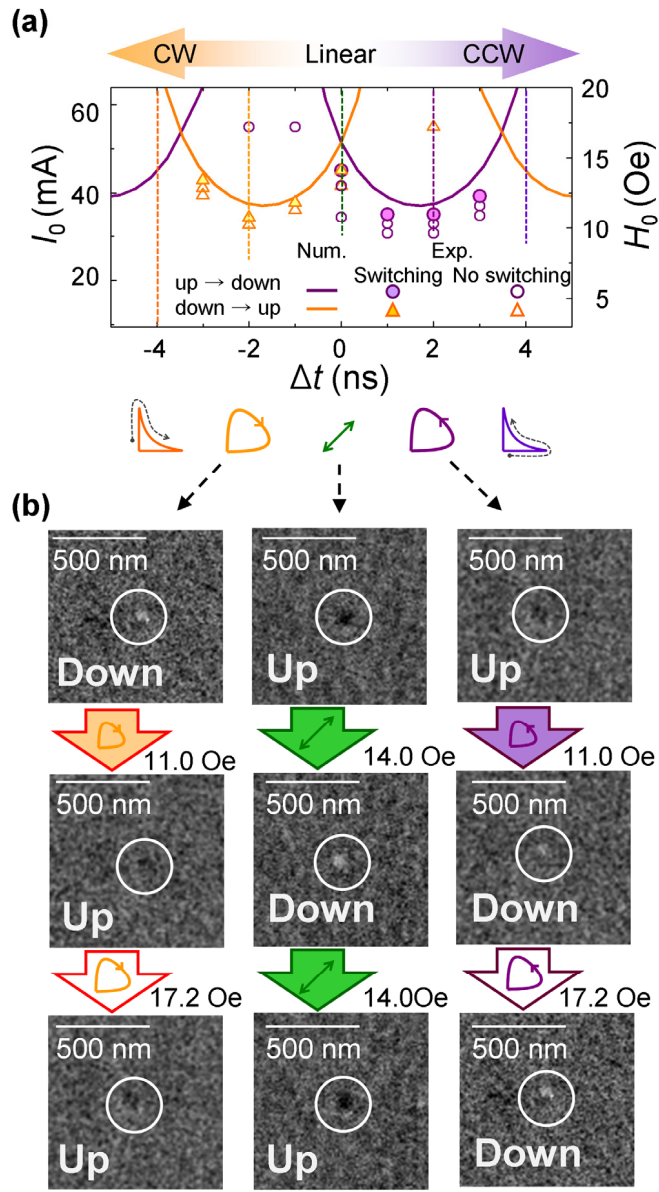

**FIG. 10**

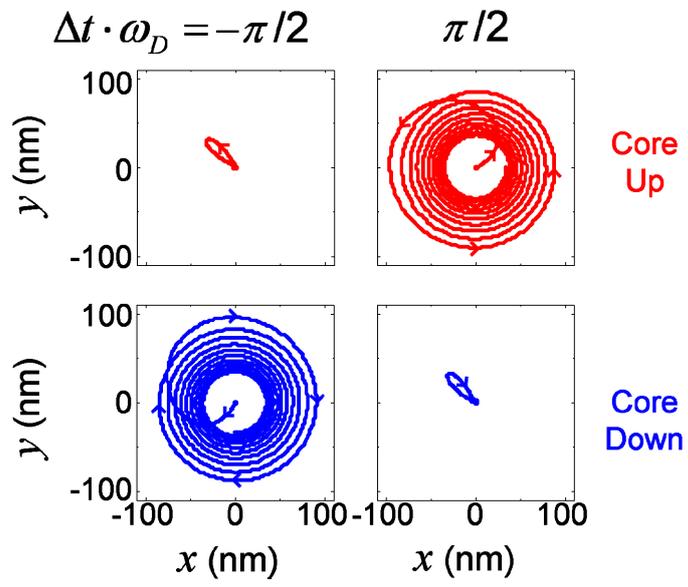